\documentclass{article}

\PassOptionsToPackage{numbers, compress}{natbib}

\usepackage[preprint]{neurips_2023}

\usepackage[utf8]{inputenc} 
\usepackage[T1]{fontenc}    
\usepackage{hyperref}       
\usepackage{url}            
\usepackage{booktabs}       
\usepackage{amsfonts}       
\usepackage{nicefrac}       
\usepackage{microtype}      
\usepackage{xcolor}         
\usepackage{graphicx}
\usepackage{amsmath}
\usepackage{subfigure}
\usepackage{algorithmic}
\usepackage[ruled,linesnumbered]{algorithm2e}
\usepackage{wrapfig}
\usepackage{color} 

\usepackage{multirow}
\usepackage[most]{tcolorbox}
\usepackage{subfigure}
\usepackage{subcaption}

\title{Verco: Learning Coordinated Verbal Communication for Multi-agent Reinforcement Learning}

\author{%
  Dapeng Li$^{1,2}$, Hang Dong$^3$, Lu Wang$^3$, Bo Qiao$^3$, Si Qin$^3$, Qingwei Lin$^3$, \and \textbf{Dongmei Zhang$^3$, Qi Zhang$^3$, Zhiwei Xu$^{1,2}$, Bin Zhang$^{1,2}$, Guoliang Fan$^{1,2}$}\\
  $^1$Institute of Automation, Chinese Academy of Sciences\\
  $^2$School of Artificial Intelligence, University of Chinese Academy of Sciences\\
  $^3$ Microsoft \\
  \texttt{lidapeng2020@ia.ac.cn} \\
}

\begin{document}

\maketitle

\begin{abstract}
In recent years, multi-agent reinforcement learning algorithms have made significant advancements in diverse gaming environments, leading to increased interest in the broader application of such techniques. To address the prevalent challenge of partial observability, communication-based algorithms have improved cooperative performance through the sharing of numerical embedding between agents. However, the understanding of the formation of collaborative mechanisms is still very limited, making designing a human-understandable communication mechanism a valuable problem to address. In this paper, we propose a novel multi-agent reinforcement learning algorithm that embeds large language models into agents, endowing them with the ability to generate human-understandable verbal communication. The entire framework has a message module and an action module. The message module is responsible for generating and sending verbal messages to other agents, effectively enhancing information sharing among agents. To further enhance the message module, we employ a teacher model to generate message labels from the global view and update the student model through Supervised Fine-Tuning (SFT). The action module receives messages from other agents and selects actions based on current local observations and received messages. Experiments conducted on the Overcooked game demonstrate our method significantly enhances the learning efficiency and performance of existing methods, while also providing an interpretable tool for humans to understand the process of multi-agent cooperation. 
\end{abstract}

\section{Introduction}
Cooperative multi-agent reinforcement learning~(MARL) has received widespread research attention due to its extensive practical applications in fields such as traffic control\cite{arel2010reinforcement}, multi-robot control\cite{matignon2012coordinated}, and sensor networks\cite{zhang2011coordinated}. However, key issues such as sample efficiency\cite{yarats2021improving, adaps}, non-stationarity\cite{papoudakis2019dealing}, and interpretability\cite{milani2022maviper, interpretableRL} of cooperation have become obstacles to further advancing these practical applications. In order to address these challenges, significant progress has been made in cooperative MARL. Among them, the centralized training and decentralized execution paradigm\cite{vdn,qmix,coma,mappo,maddpg,sea} alleviate the non-stationary problem during multi-agent training by introducing additional information during the training period. However, due to partial observability, the strategies learned by agents may be fragile since the uncertainty of other agents during execution may lead to catastrophic incoordination and sub-optimality~\cite{miller2002communication,yang2020overview}. 
Inspired by human cooperation, a series of works achieve communication between individuals by exchanging their observations or hidden embedding\cite{li2023explicit,bicnet,commnet}, thereby stabilizing the learning process and promoting more efficient cooperation among agents. These methods usually treat messages as black boxes, send numerical messages, and assume that the policy network can adaptively extract content that is helpful for learning during the learning period. Therefore, the communication content is usually presented in a form that humans cannot understand, making it hard to interpret the communication mechanism.

\begin{wrapfigure}{r}{0.6\textwidth}
\vspace{-20pt}
\begin{center}
\includegraphics[width=0.6\textwidth]{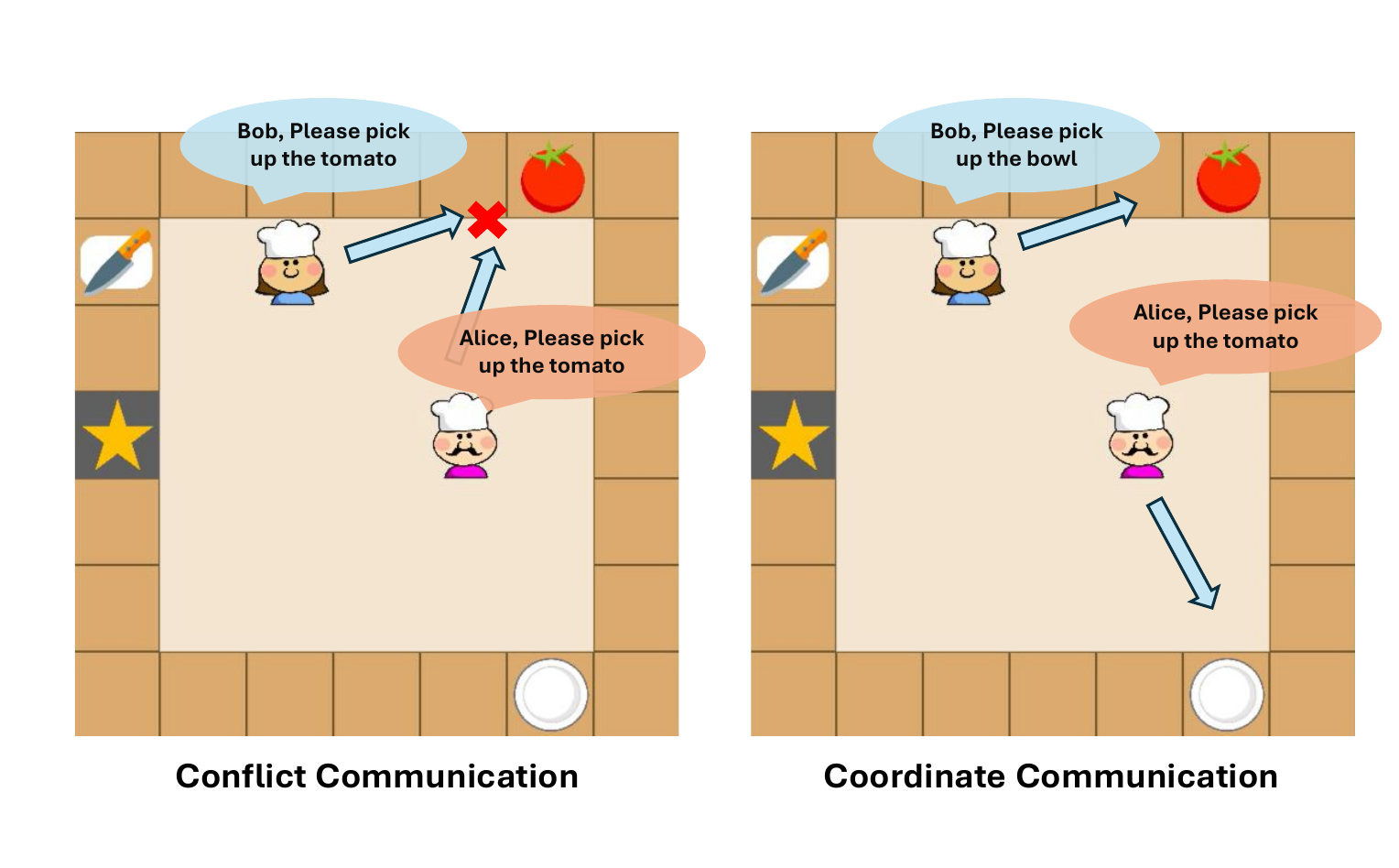}
\end{center}
\vspace{-15pt}
\caption{Incorrect messages can easily lead to conflicts, while coordinated messages can promote efficient cooperation among agents. }
\vspace{-10pt}
\label{conflict}
\end{wrapfigure}

One natural and interpretable way of communication is to directly generate verbal language as communication messages, which also means the policy network needs to have the ability to understand verbal text. Recent work has shown that using large language models for understanding knowledge texts can effectively improve sample efficiency in complex decision-making tasks\cite{tan2024true, carta2023grounding}. By aligning prior knowledge of large language models (LLMs) with the functional requirements of the environment with only a small amount of environmental interaction data, the LLM can achieve good performance. Meanwhile, as LLMs use verbal text as model input, it is natural to use LLMs for verbal communication.  However, there are key difficulties in directly generating verbal messages: \romannumeral1) The candidate space for generating messages is too large to explore. \romannumeral2) The text output lacks gradients and cannot be optimized end-to-end with environmental rewards. \romannumeral3) The messages generated by agents based on local observations are prone to conflicts (as shown in Figure~\ref{conflict}). 
 
To address the aforementioned challenges, this paper proposes a novel multi-agent algorithm for learning coordinated \textbf{VER}bal \textbf{CO}mmunication~(\textbf{Verco}) that is understandable to humans. We first use a powerful LLM (e.g., GPT-4) as the teacher model and perform Supervised Fine-tuning (SFT) on the student model equipped with low-rank adapters (LoRA)~\cite{hu2021lora} based on the output of the teacher model. The policy learning module is then fine-tuned online through interaction with the environment, aligning the model's prior knowledge with the environment. We train different LoRA parameters for the communication module and the policy learning module respectively, avoiding mutual interference between the two modules and the cost problem of managing and training multiple large models. Overall, the communication module generates and sends text messages based on the local observations of the agents to promote cooperation between them, while the action module receives local observations and messages sent by other agents and selects the action to be taken from the candidate action set.
We have conducted experiments in different scenarios on the popular multi-agent decision-making environment Overcooked and compared Verco with existing baseline methods. We find that introducing verbal communication can significantly improve the performance of agents, and also improve the interpretability of cooperation. 

\section{Related Work}
In this section, we will introduce some of the work related to our paper.
\subsection{LLMs for Decision Making}
LLMs trained on large datasets have shown remarkable abilities in various downstream tasks. Recent works have gradually applied LLMs to complex tasks such as robot control~\cite{roco, jiang2023vima}, planning generation~\cite{liu2023llm+}, embodied agent~\cite{saycan, zhang2023proagent, zhang2023controlling, wang2023voyager}, etc. Among them, SayCan~\cite{saycan} and LLM+P~\cite{liu2023llm+} use the LLM as a high-level planner to generate long-term plans for agents based on task goals but often do not directly interact with the environment. Voyager~\cite{wang2023voyager} uses GPT-4 to accomplish highly complex tasks in the Minecraft game and continuously generates new skills during the learning process. ReAct~\cite{yao2022react} combines reasoning and action to enhance LLMs' reasoning ability for high-level goals, while improving the interpretability and credibility of LLMs' decisions. As a supplement, \cite{dasgupta2023collaborating} added additional reporters to provide useful information for the planner. However, these methods cannot be improved based on environmental feedback, making it difficult to adapt to different environments. In recent works, GLAM~\cite{carta2023grounding} and TWOSOME~\cite{tan2024true} both use reinforcement learning (RL) to enable LLMs to interact and learn in a single-agent environment, and the rich prior experience of LLMs itself can significantly reduce the learning cost of agents. 

\subsection{Finetuning LLMs}
In recent works, finetuning has been employed to improve the performance of LLMs in specific tasks across different domains. Among these approaches, the use of RL to enhance the consistency between the LLM and human preferences is a common practice~\cite{ouyang2022training}. Many reinforcement learning with human feedback (RLHF) methods utilize PPO~\cite{schulman2017proximal} to learn a reward function from human datasets and fine-tune LLMs according to the learned reward function. A significant challenge in fine-tuning LLMs is how to reduce training costs. Parameter-efficient finetuning (PEFT)\footnote{\href{https://github.com/huggingface/peft}{https://github.com/huggingface/peft}.}~\cite{peft,ding2023parameter} can significantly reduce the number of LLM parameters while avoiding excessive performance loss. As a recent technology in PEFT, low-rank adapters~(LoRA) indirectly train the dense layers of neural networks by learning low-rank matrices. TWOSOME~\cite{tan2024true} fine-tuned a LoRA as the actor model, thus enabling efficient fine-tuning while avoiding mutual interference between actor and critic.

\subsection{MARL with Communication}
Existing multi-agent cooperation algorithms have made significant progress in many complex scenarios, but the realistic situation of partial observability greatly limits the degree of cooperation among agents. To address this issue, researchers have proposed a series of communication-based cooperation paradigms~\cite{miller2002communication, bicnet, commnet, maic, li2023explicit} to facilitate agents' understanding of the environment and teammates by exchanging local observations or intentions. However, most of the communication messages generated by existing communication algorithms are difficult for humans to understand, thus not interpretable and cannot be improved explicitly. To make the communication messages understandable and ineterpretable, using verbal text as a communication message naturally becomes a solution. The FAMA~\cite{slumbers2023leveraging} algorithm extends GLAM to multi-agent settings and introduces a verbal communication module. However, the communication module in the FAMA method cannot be learned through interactions with the environment and can only select from a pre-defined set of message candidates, significantly reducing the degree of message generation freedom and relying on the quality of the pre-specified message candidates. 

\section{Preliminaries}

\newcommand{\tup}[1]{G} 
\newcommand{\Obs}[1]{\Omega_{{#1}}} 
\newcommand{\obs}[1]{o_{{#1}}} 
\newcommand{\obsfunc}[1]{O_{{#1}}} 

\newcommand{\Hidden}[1]{H_{{#1}}} 
\newcommand{\hidden}[1]{h_{{#1}}} %
\newcommand{\State}[1]{S_{{#1}}} 
\newcommand{\state}[1]{s_{{#1}}}
\newcommand{\statetrans}[1]{\mathcal{T}_{{#1}}}

\newcommand{\Act}[1]{U_{{#1}}}
\newcommand{\act}[1]{u_{{#1}}}
\newcommand{\enc}[1]{g_{{#1}}}
\newcommand{\Traj}[1]{\tau_{{#1}}}
\newcommand{\traj}[1]{\tau_{{#1}}}

\newcommand{\agent}[0]{i}

\newcommand{\agentn}[0]{n}
\newcommand{\Agentn}[0]{\mathcal{N}}

\newcommand{\att}[1]{v_{{#1}}}
\newcommand{\rec}[1]{\hat{v}_{{#1}}}

\newcommand{\attweight}[1]{w_{{#1}}}

\newcommand{\voc}[1]{\mathcal{V}_{{#1}}}

\newcommand{\reward}[1]{r_{{#1}}} 
\newcommand{\policy}[1]{\pi_{{#1}}} 
\newcommand{\stynum}[0]{K}

\subsection{Problem Formulation}
We assume a textual MARL setting here with language vocabulary $\mathcal{V}$. The multi-agent problem here can be described as a partially observable Markov game~(POMG), which is defined by the following tuple $\tup{} = \left< \State{}, \Act{}, \Agentn{}, \Obs{}, \statetrans{}, \obsfunc{}, \reward{},\gamma, \mathcal{V}\right>$, with $\State{}$ the state space, $\Act{}\subset\voc{}$ the action space. At each discrete time step $t$, each agent $\agent\in\Agentn{}:=\{1,\dots,\agentn{}\}$ will select an action $\act{\agent}\in \Act{\agent}\subset\voc{}$. $\statetrans{}(\state{}'|\state{},\boldsymbol{\act{})}:\State{}\times \Act{} \times \State{} \rightarrow P(\State{})$ is the state transition function, where $\boldsymbol{\act{}}=\{\act{1},\dots,\act{\agentn{}}\}\in\boldsymbol{\Act{}}\equiv\Act{}^\agentn{}$ is the joint action. Each agent $\agent$ can get its textual partial observation $\obs{\agent}\in \Obs{}$ by the observation function $\obsfunc{}(\state{}, \agent): \State{} \times \mathcal{N} \rightarrow \Obs{}$. $\reward{\agent}(\state{},\act{\agent}):\State{}\times \Act{\agent} \rightarrow \mathbb{R}$ is the reward function for each agent $\agent$. The $\gamma$ is the discount factor. The goal for each agent is to maximize the expected return. In order to meet the demands of communication and decision-making, we designed a dual-LLM structure for agents, enabling them to send text messages based on their local observations while making decisions based on the received messages from other agents. By considering both local observations and teammates' messages, the agents can select corresponding coordinated actions. Our entire framework consists of three main components: the teacher policy $\pi^{tch}$, the message policy $\pi_{\eta}$, and the action policy $\pi_{\theta}$.

\subsection{LLM and Finetuning} 
We consider using LLMs as interactive agents in embodied environments, the LLMs are trained on massive amounts of text data and can generate human-like responses to questions or prompts. In order to improve the performance of LLMs in various specific tasks, fine-tuning is commonly employed. Among these, LoRA (Low-Rank Adaptation)~\cite{hu2021lora} serves as a popular fine-tuning method, effectively reducing computational costs while maintaining satisfactory performance. Specifically, LoRA training the dense layers with weights matrix, $W_0\in \mathbb{R}^{d\times k}$ and injects trainable rank decomposition matrices into the Transformer by $W_0 + \Delta W = W_0 +BA$, where $B\in\mathbb{R}^{d\times r}, A\in\mathbb{R}^{r\times k}$, since the rank $r$ is much smaller than $d$ and $k$, it can significantly reduce the scale of trainable parameters.

\section{Method}
\begin{figure*}[t]
    \centering
    \includegraphics[width=5.5 in]{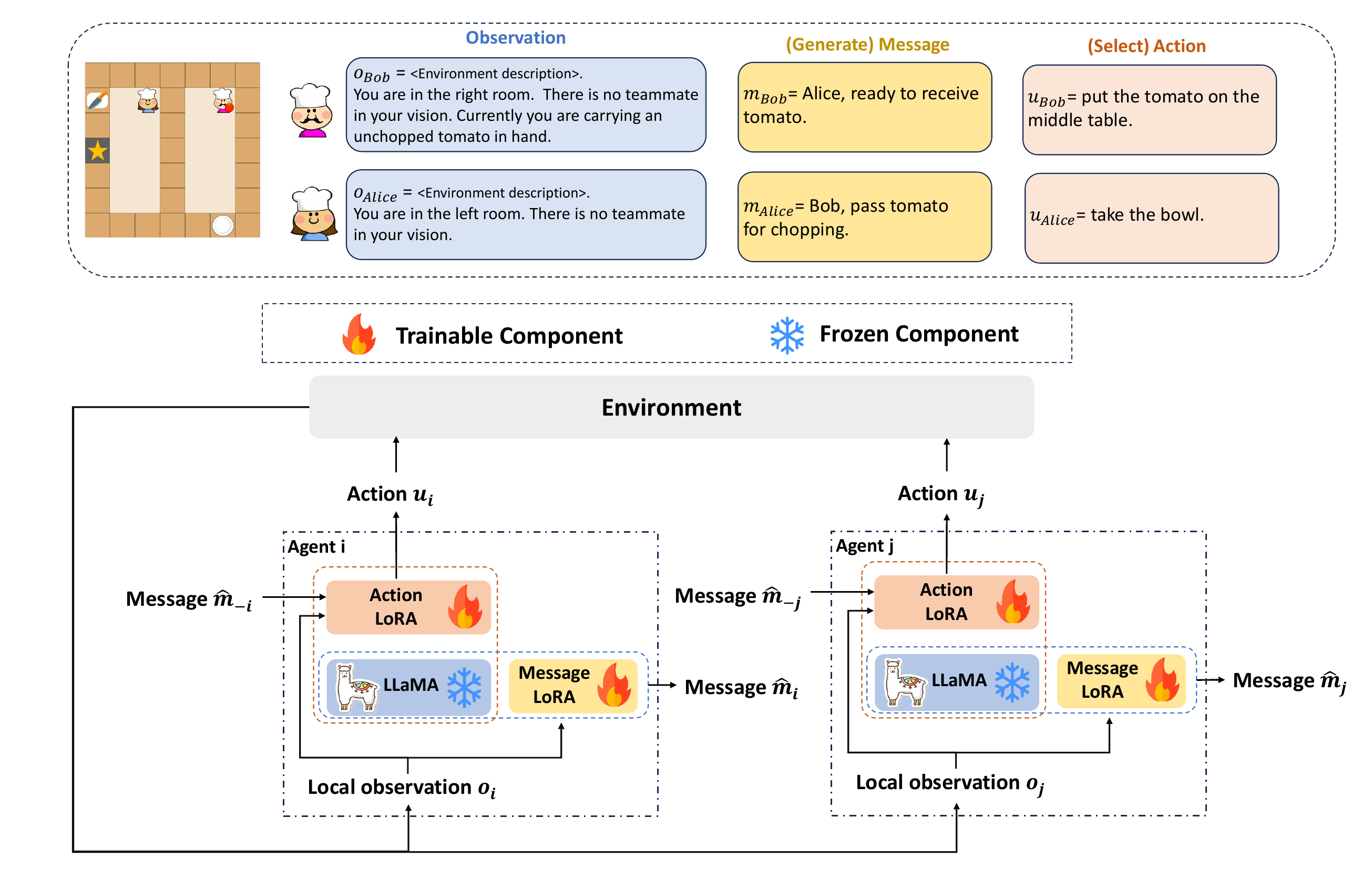}
    \caption{Verco framework: We first finetune the LoRA weight of the communication module with the global message label. Then we load the LoRA weight so the agent can directly generate verbal messages with its local observation. Meanwhile, the action policy takes the current local observation and text messages from other agents as input and outputs the decision. The action policy fine-tunes the weights using PPO based on the rewards returned by the environment.}
    \label{fig1}
\end{figure*}

We introduce the LLM, capable of text generation, into agents as a communication module and action module to enhance the collaborative capabilities and interpretability of multi-agent systems. To accelerate training and improve the quality of communication messages, we first generate communication labels for supervised training of the communication module. Subsequently, the fine-tuned communication module weights are loaded into the agents, allowing them to generate highly flexible text messages to facilitate coordination among agents. The entire flow is shown in Figure~\ref{fig1}
The action modules will interact and improve within the environment using RL. In this section, we will provide a detailed introduction to each part of the entire algorithmic. 

\subsection{Cooperation with verbal communication}
In the context of human collaborative problems, the diversity in the information received by individuals and their distinct modes of thinking often render simple independent actions insufficient for achieving coordinated consensus. Consequently, the use of natural language has become a prevalent and significant method for facilitating coordination and cooperation among humans. From another perspective, the domain of large language models is precisely dedicated to and adept at enhancing communication and interaction in a human-like manner. Therefore, it is a natural progression to integrate large language models into agents as communication modules. These agents can employ the communication module to convey information or intentions observed by themselves to others, thereby enabling more effective collaboration.

\begin{figure*}[t]
    \centering
    \includegraphics[width=5.5 in]{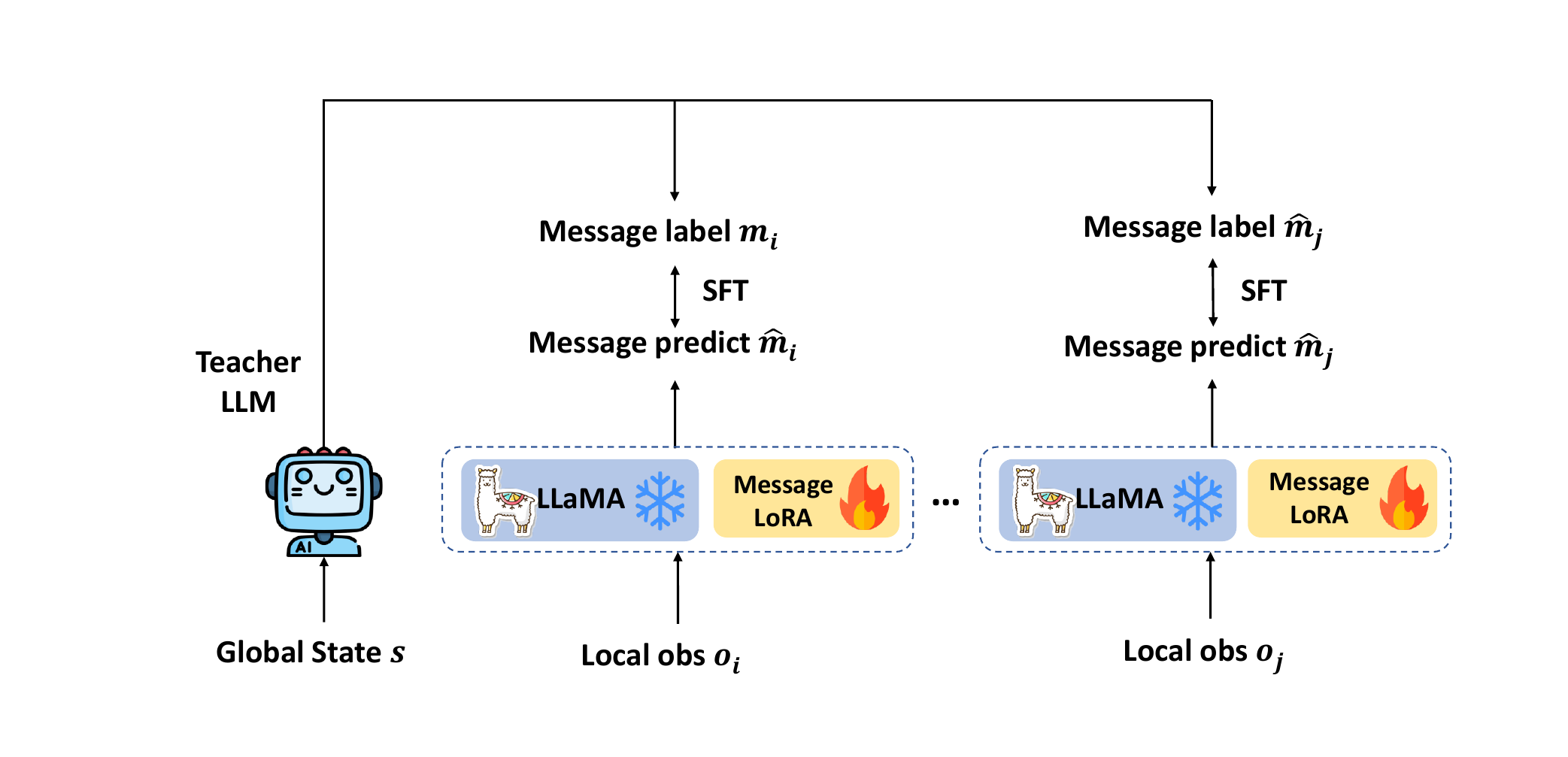}
    \caption{Message module SFT stage: We employ a large model (GPT-4) as the teacher model to generate message samples based on global observations, and distill the learning for a smaller language model (LLaMA-7B) as the communication model $\pi_\eta$.}
    \label{fig2}
\end{figure*}

\subsection{Coordination Message Policy Pre-training} 
Relying solely on environment interaction to train a communication model that automatically generates reasonable messages from scratch is highly inefficient, while utilizing a large-scale model (e.g., GPT-4, LLaMA-65B) for communication is both costly and difficult to fine-tune. To facilitate the generation of coordinated and consistent messages in the communication module, we employ GPT-4 as a coordinated message generator, taking the local observations of all agents as input and generating customized message labels for each agent $\{m_i\}^n_{i=1} \sim \pi^{tch}(p^{tch})$. The corresponding prompt $p^{tch} = \rho_{tch}(\{o_i\}^n_{i=1}, n)$ is obtained through the teacher LLM's prompt function $\rho_{tch}$:

\begin{tcolorbox}[title=Teacher LLM prompt,
                  colback=gray!10,
                  colframe=black,
                  width=13.8cm,
                  arc=3mm, auto outer arc,
                 ]
  You are a message design assistant who needs to design messages for <$n$> agents, <\textit{Brief environment description}>.\\
  The observation of Agent $1$: <{$o_{1}$}>, \\
  ... \\
  The observation of Agent $n$: <{$o_{n}$}>.\\
  To complete <\textit{Task Goal}>, design concise messages (within 10 words) for each agent to their teammate: <\textit{LLM response here}>
\end{tcolorbox}

Supervised learning training is conducted on the message samples using LLaMA-7B, which is updated by LoRA. The message prediction of agent $i$ is generated by message policy $\hat{m}_{i} \sim \pi_{\eta}(\cdot|\rho_{m}(o_{i}))$. The SFT loss can be expressed as follows: 
\begin{equation}
\mathcal{L}_{SFT} = \sum^n_{i=1}{CROSS\_ENTROPY(\hat{m}_i, m_i)},
\label{sft}
\end{equation}
where $CROSS\_ENTROPY$ represents the cross entropy loss function.
The entire workflow of SFT is shown in Figure~\ref{fig2}(a)
To control the training scale, we let all agents share the action policy and message module.  Besides, We use initialized agents to interact in the environment for several rounds to collect trajectory data. Empirically, with only a small amount of data and training steps, the communication policy $\pi_{\eta}$ can already generate a reasonably coherent verbal message.

\subsection{Action Policy Alignment}
The action module also employs LLMs, which brings two distinct advantages:  \romannumeral1) LLMs can directly and effectively comprehend verbal messages from teammates without the need for further training. \romannumeral2) LLMs have been demonstrated to have learned a substantial amount of physical rules present in the real world~\cite{patel2021mapping}, and this knowledge can significantly reduce the number of interactions required between the policy and the environment, thereby enhancing sample efficiency. To align general purpose LLMs with the specific environment, feedback from the environment is utilized to adapt to the environment better and yield better performance.

Inspired by previous works~\cite{carta2023grounding, tan2024true}, we do not directly prompt the LLM to generate the actions, instead, we query the LLM for the probabilities of all available actions. Assume the $k$-th action of $i$-th agent $u_{i,k}\in U_i$ is a sequence of tokens $u_{i,k} = \{w^1_{i,k},\dots,w^{N_{i,k}}_{i,k}\}$, where $N_k$ is the token number of the $k$-th action. So the token-level probability of $u_{i,k}$ can be calculated by: 
\begin{equation}
    P_{token}(u_{i,k}|\rho_u(o_i),\hat{m}_{-i})=\prod^{N_{i,k}}_{j=0}P(w^j_{i,k}|\rho_u(o_i),\hat{m}_{-i},w^{<j}_{i,k}),
\end{equation}
where $\rho_u$ is the action prompt function. Therefore, the action policy $\pi$ can be got by the softmax over the token-level probabilities over actions:
\begin{equation}
    P(u_{i,k}|\rho_{u}(o_i),\hat{m}_{-i})=\frac{\text{exp}(\text{log}P_{token}(u_{i,k}|\rho_u(o_i),\hat{m}_{-i}))}{\sum_{u_i\in U_i}\text{exp}(\text{log}P_{token}(u_i|\rho_u(o_i),\hat{m}_{-i}))}.
\end{equation}

In our work, we employ PPO~\cite{de2020independent} to optimize the action policy. PPO is a state-of-the-art actor-critic RL method, where each agent learns an actor parameterized by $\theta$ and a critic parameterized by $\phi$.

For each agent $i$, the policy loss can be formulated as follows:
\begin{equation}
\begin{aligned}
\mathcal{L}_i(\theta) = &\mathbb{E}_{o^t_{i},u^t_{i}}[\text{min}(\frac{\pi_{\theta}(u^t_{i}|\rho_u({o}^t_{i}),\hat{m}^t_{-i})}{\pi_{\theta_{old}}(u^t_{i}|\rho_u({o}^t_{i}),\hat{m}^t_{-i})}A^t_{i},\\
&\text{clip}(\frac{\pi_{\theta}(u^t_{i}|\rho_u({o}^t_{i}),\hat{m}^t_{-i})}{\pi_{\theta_{old}}(u^t_{i}|\rho_u({o}^t_{i}),\hat{m}^t_{-i})},1-\epsilon,1+\epsilon)A^t_{i})],
\label{actor}
\end{aligned}
\end{equation}

where $A^t_i$ is the Generalized Advantage Estimation~(GAE)~\cite{schulman2015high}.
The critic network is composed of an additional MLP added to the last transformer block of the LLaMA model. 

\begin{equation}
\begin{split}
\mathcal{L}_i(\phi) = &\mathbb{E}_{o^t_i}[\text{min}\{(V_{\phi}(\rho_u(o^t_i)) - \hat{V^t_i})^2, (V_{\phi_{old}}(\rho_u(o^t_i)) + \\
 & \text{clip}(V_\phi(\rho_u(o^t_i))-V_{\phi_{old}}(\rho_u(o^t_i)),-\epsilon, +\epsilon) - \hat{V}^t_i)^2\}],
\end{split}
\label{critic}
\end{equation}
where $\hat{V}^t_i = A^t_i + V_{\phi_{old}}(\rho_u(o^t_i))$.

The overall learning loss is:
\begin{equation}
    \mathcal{L}_{RL} = \sum^{n}_{i=1}\mathcal{L}_i(\theta) + \lambda_{critic}\mathcal{L}_i(\phi)+\lambda_{entropy}\mathcal{H}(\pi_{i,\theta}),
\label{overall}
\end{equation}
where $\mathcal{H}(\pi_{i,\theta})$ denotes the entropy of agent $i$'s action policy $\pi_{i,\theta}$.

Given that the goals of the action policy and message policy are distinct, we load two independent LoRA weights separately to ensure that they do not interfere with each other. During the entire action policy alignment stage, the message policy will be frozen to stabilize the training of the action policy. The actor, critic, and message networks share the same LLaMA model and the gradients do not affect each other, therefore greatly reducing our training costs and memory requirements. The entire algorithm can be described as shown in Algorithm~\ref{algo1}.
\begin{algorithm}[ht]
\caption{Training Procedure for Verco}
\SetKwData{Left}{left}\SetKwData{This}{this}\SetKwData{Up}{up}
\SetKwFunction{Union}{Union}\SetKwFunction{FindCompress}{FindCompress}
\SetKwInOut{Input}{Input}\SetKwInOut{Output}{Output}
\Input{initialise action policy parameters $\theta$, value function parameters $\phi$, and message policy parameters $\eta$. Set the data buffer for SFT $D_{SFT} = \emptyset$ and for RL training $D_{RL} = \emptyset$}
\Output{$\theta^*$, $\phi^*$, and $\eta^*$}
\begin{algorithmic}[1] 
\label{algo1}
\FOR{k = 1,2,\dots,$K$}
\STATE \#Collect trajectory with initial action policy $\pi_\theta$.
\FOR{each time step $t$}
\STATE Generate message label given global state: $\{m^t_i\}^n_{i=1}\sim\pi^{tch}(p^{tch})$.
\FOR{i = 1,2,\dots,$n$}

\STATE Generate action from initial action policy $u^t_i\sim\pi_\theta(\rho_u(o^t_i))$.
\STATE Add message label and observation to buffer:$D_{SFT} = D_{SFT}~\cup~(m^t_i,o^t_i)$
\ENDFOR
\STATE Take joint actions $\boldsymbol{u}^t$ and obtain the next observation $\boldsymbol{o^{t+1}}$.
\ENDFOR
\ENDFOR

\FOR{each batch numbers}
\STATE Sample a batch of data from data buffer $D_{SFT}$
\STATE Update the message policy $\pi_{\eta}$ following Eq. \ref{sft}.
\ENDFOR

\FOR{each episode}
\STATE \#RL training stage with frozen message policy $\pi_\eta$.
\FOR{each $t$}
\FOR{i = 1,2,\dots,n}
\STATE Generate messages for each agent from the message policy $\hat{m}^t_i\sim\pi_\eta(\cdot|\rho_m{(o^t_i)})$.
\ENDFOR
\FOR{i = 1,2,\dots,n}
\STATE Generate actions for each agent from action policy with received message ${u}^t_i\sim\pi_\theta(\cdot|\rho_u{(o^t_i)},m^t_{-i})$.
\ENDFOR
\STATE Take joint actions $\boldsymbol{u}^t$ and obtain the next observation $\boldsymbol{o^{t+1}}$.
\STATE Store the transition to buffer: $D_{RL}=D_{RL}~\cup~(\boldsymbol{o}^t,\boldsymbol{u}^t,\boldsymbol{r}^t,\boldsymbol{o}^{t+1},\boldsymbol{m}^t)$.
\ENDFOR
\STATE Update the policy parameter $\theta$ and value function parameter $\phi$ by minimize the overall learning loss in Eq.~\eqref{overall}.
\ENDFOR
\end{algorithmic}
\end{algorithm}

\begin{figure*}[t]
\subfigure[Single Room]{
\begin{minipage}{0.32\linewidth}
\centerline{\includegraphics[width=0.8\textwidth]{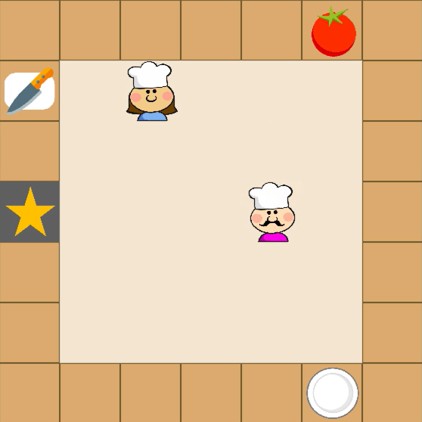}}
\label{fig:mapA}
\end{minipage}}
\subfigure[Separate Rooms]{
\begin{minipage}{0.32\linewidth}
\centerline{\includegraphics[width=0.8\textwidth]{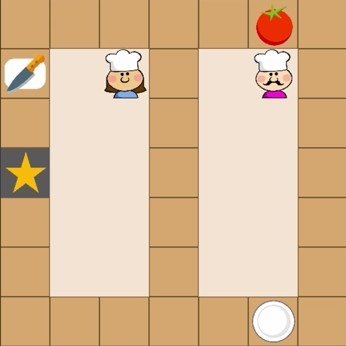}}
\label{fig:mapB}
\end{minipage}}
\subfigure[Cooking process]{
\begin{minipage}{0.32\linewidth}
\centerline{\includegraphics[width=0.8\textwidth]{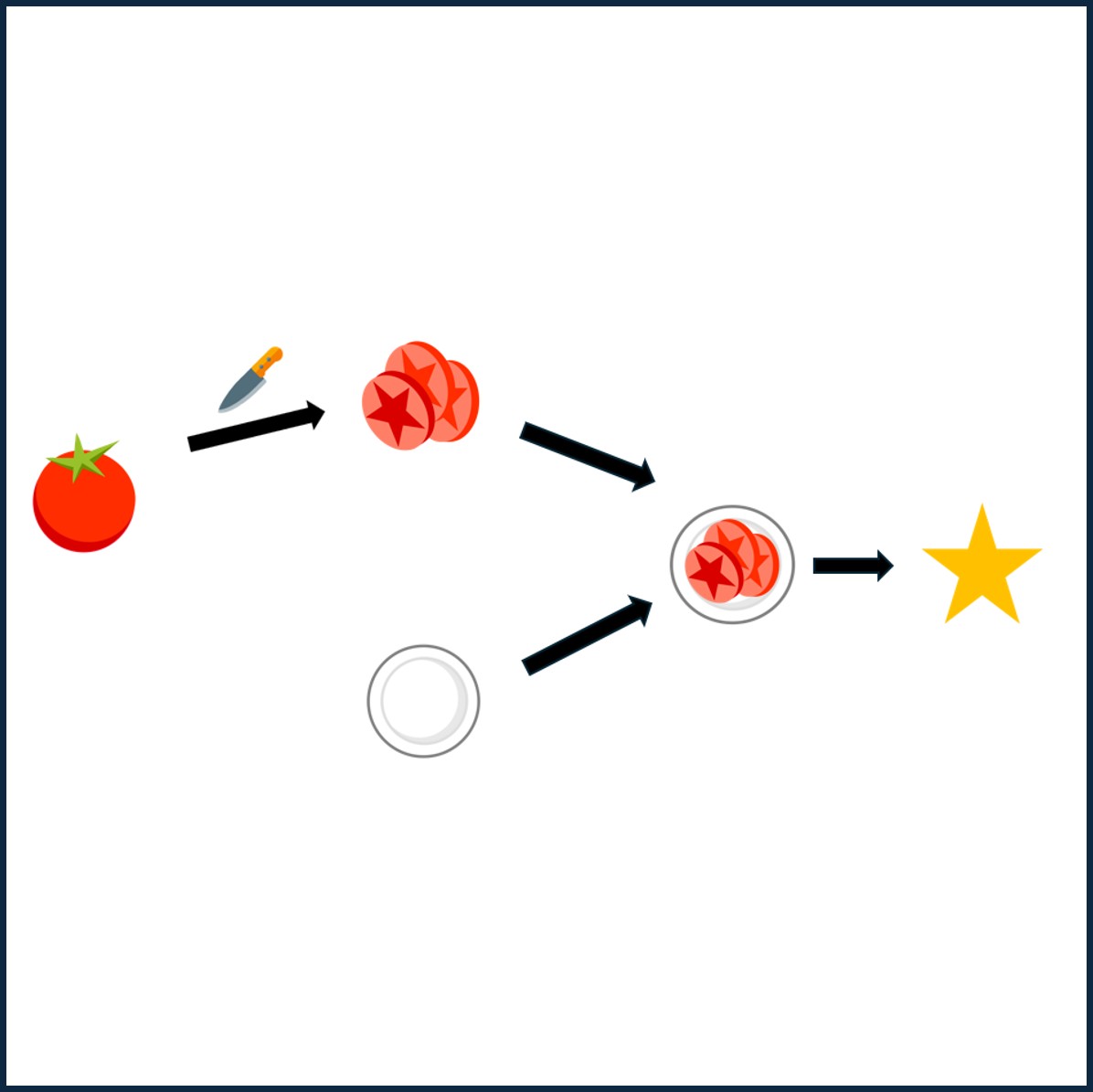}}
\label{fig:makedish}
\end{minipage}}
\caption{Experimental environments. Figure~\ref{fig:mapA} and Figure~\ref{fig:mapB} show two different maps in Overcooked. Figure~\ref{fig:makedish} shows the production process of tomato salad.}
\label{fig:env_descript}
\end{figure*}
\section{Experiments}
In this section, we first introduce the experimental scenario. Then, we provide a detailed introduction to all baselines. We demonstrate the performance of baseline algorithms in different experimental scenarios. We also visualize the communication content of the Verco algorithm. 

\subsection{Environment description} The Overcooked environment is a popular complex environment for decision-making problems. The goal for agents is to make different types of salads with the provided raw materials and tools in a 7x7 grid-size kitchen. Our work extends the single-agent textual version of Overcooked in \citeauthor{tan2024true} to multi-agent systems. In our setting, two agents need to collaborate to complete tasks in the environment. As shown in the figure, there are two types of maps A and B. In Map A, two agents are in the same space, so there may be collisions. In Map B, the two agents are separated and need to complete the task by passing items through the table. The environment is partially observable, and each agent can only observe the objects within 5×5 square centered of itself. As to the reward, chopping the correct item will be +0.2, providing the correct dish will be +1, -0.1 for delivering any wrong item, -0.01 for each collision between agents, and -0.001 for each time step. 

\subsection{Baselines}

We compare our method with two trainable baselines and a baseline for direct decision-making by GPT-4. The description for each baseline is as follows:

\textbf{TWOSOME~\cite{tan2024true}.} TWOSOME proposes an efficient single-agent LLM decision fine-tuning framework, and it balances the joint probability of candidate actions by several regularizing methods. 

\textbf{Symbolic PPO~\cite{schulman2017proximal,de2020independent}.} The symbolic PPO takes the raw numerical observation as its input and uses MLPs as the backbone network. 

\paragraph{CommNet~\cite{commnet}.} CommNet, as a commonly used multi-agent communication algorithm, takes the average of numerical observations from all agents and sends it as a message to each agent.

\textbf{GPT-4~\cite{openai2024gpt4}.} GPT-4 has shown great potential in decision-making problems, but a significant drawback is that GPT-4 is difficult to improve from the environment's feedback. We directly input the text context and candidate actions into GPT-4, allowing GPT-4 to select the currently appropriate action from the candidate actions.

\textbf{Verco.} Our method uses two LoRA weights for communication and action policy, respectively. The communication policy is obtained by SFT with message labels generated by GPT-4, and the action policy is learned by RL through interaction with the environment.

All curves are presented with average performance and 25$\sim$75\% deviation over four random seeds, with the solid lines representing the median win rates. Due to our efficient design, our algorithm and all experiments can be completed on a single NVIDIA Tesla V100 32GB GPU.

\begin{figure*}[ht]
\centering
\subfigure{
\begin{minipage}{0.32\linewidth}
\centerline{\includegraphics[width=1.1\textwidth]{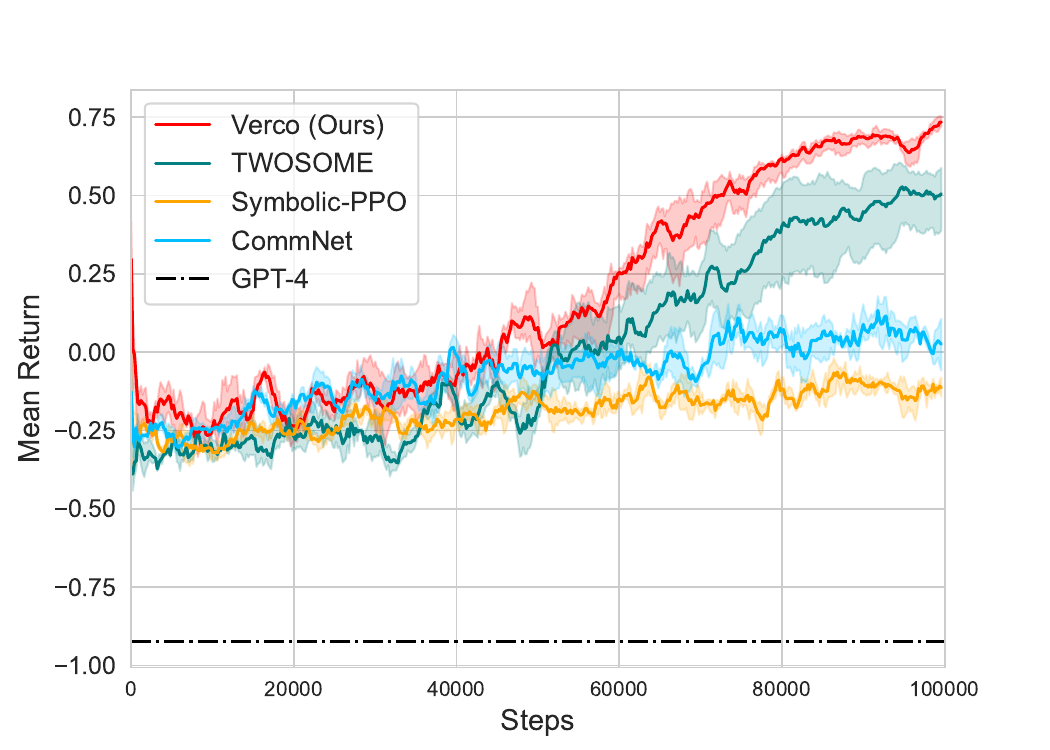}}
\label{fig:mapA_return}
\end{minipage}}
\subfigure{
\begin{minipage}{0.32\linewidth}
\centerline{\includegraphics[width=1.1\textwidth]{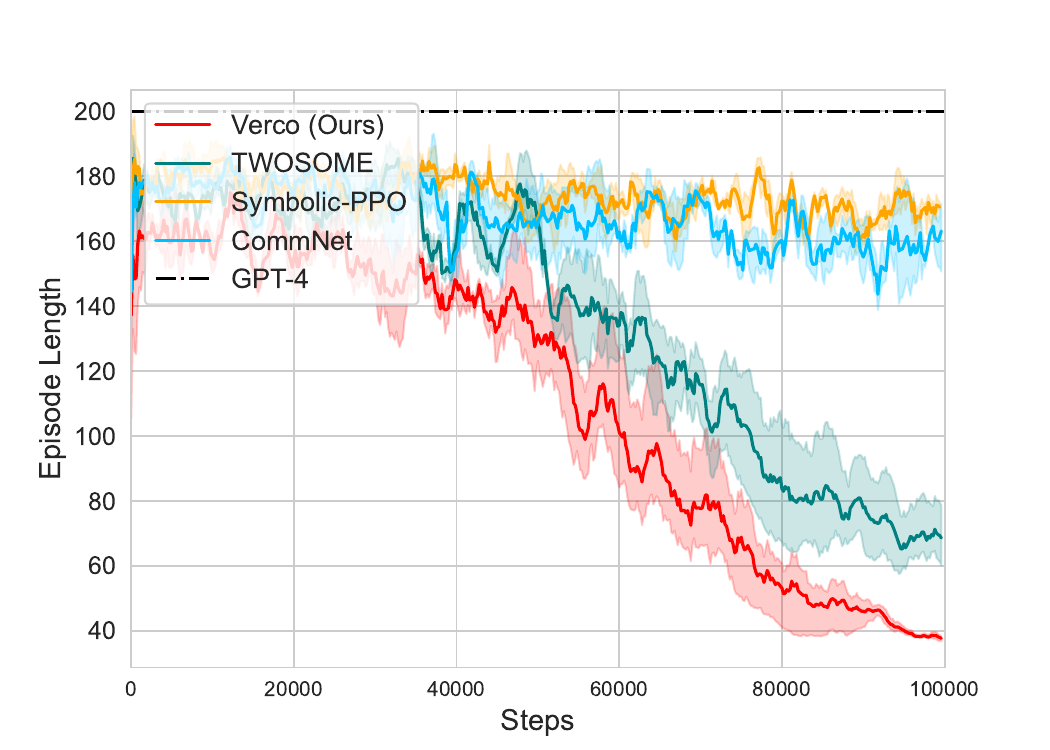}}
\label{fig:mapA_length}
\end{minipage}}
\subfigure{
\begin{minipage}{0.32\linewidth}
\centerline{\includegraphics[width=1.1\textwidth]{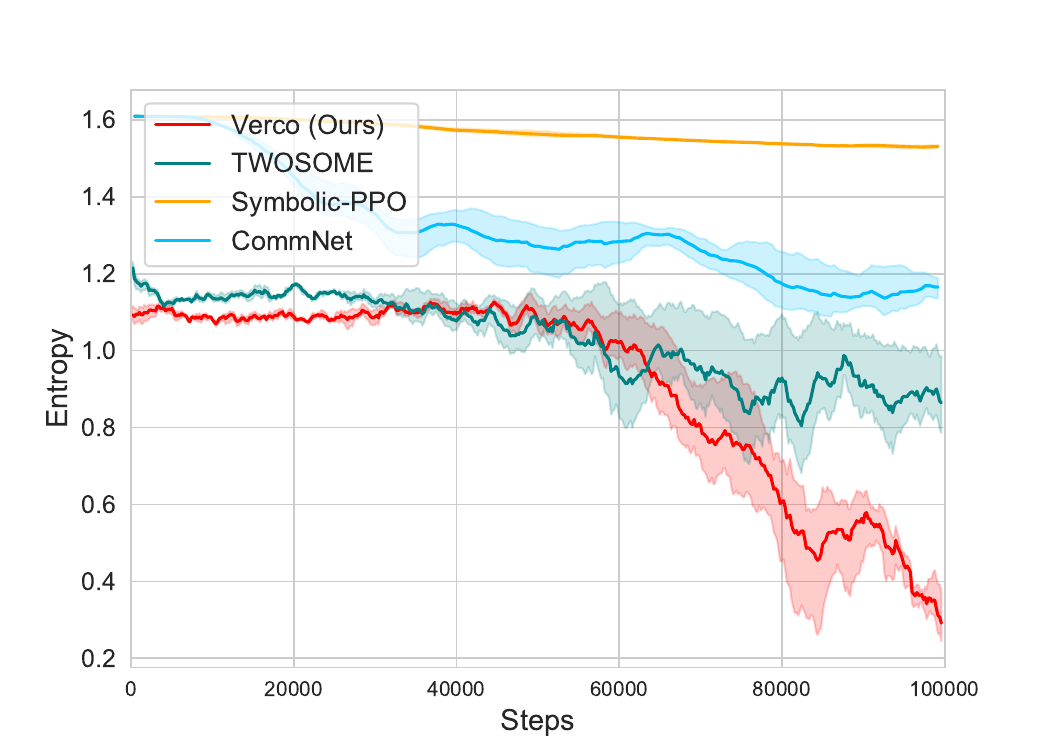}}
\label{fig:mapA_entropy}
\end{minipage}}
\centerline{(a)~Comparison of performance on Single Room}
\\
\subfigure{
\begin{minipage}{0.32\linewidth}
\centerline{\includegraphics[width=1.1\textwidth]{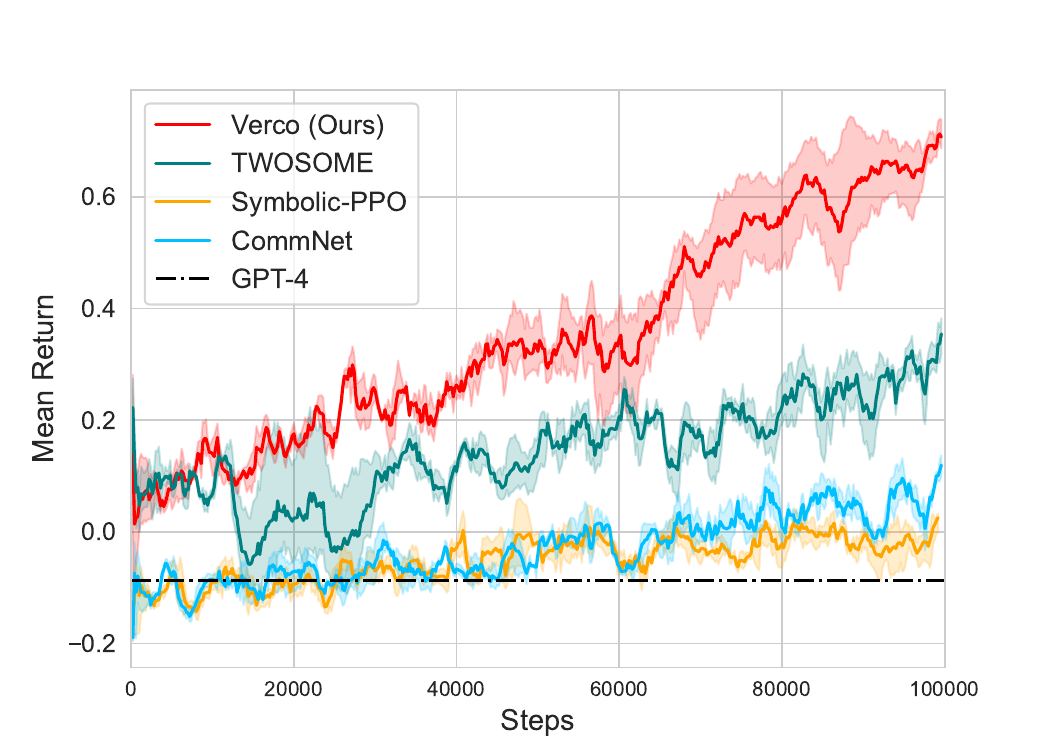}}
\label{fig:mapB_return}
\end{minipage}}
\subfigure{
\begin{minipage}{0.32\linewidth}
\centerline{\includegraphics[width=1.1\textwidth]{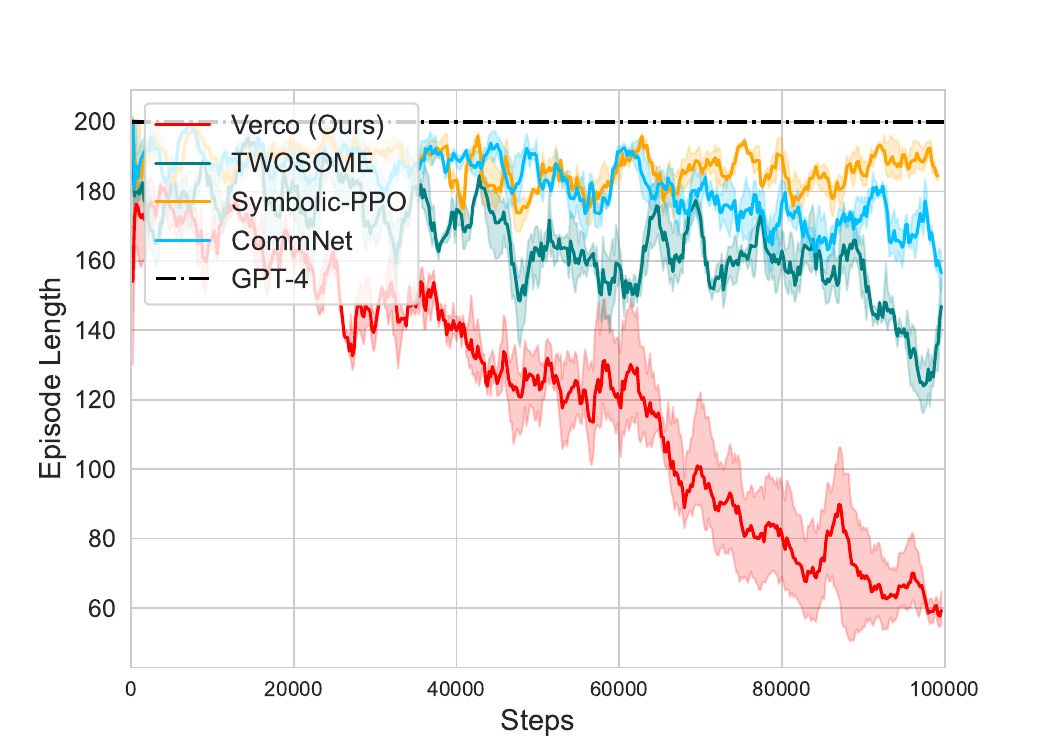}}
\label{fig:mapB_length}
\end{minipage}}
\subfigure{
\begin{minipage}{0.32\linewidth}
\centerline{\includegraphics[width=1.1\textwidth]{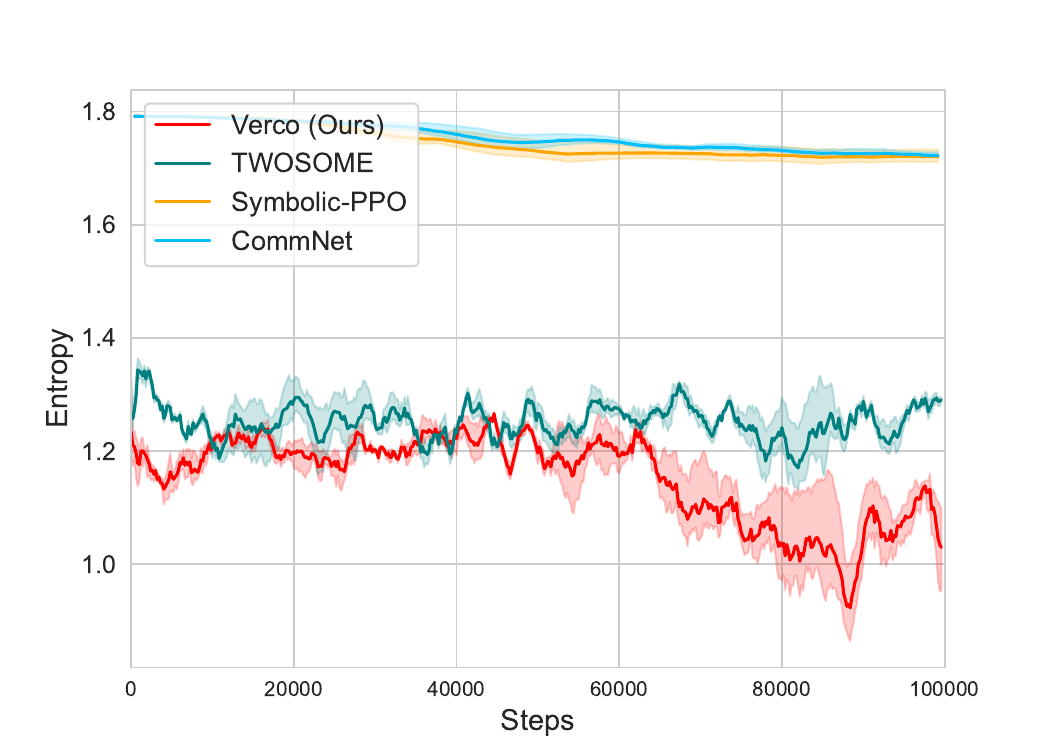}}
\label{fig:mapB_entropy}
\end{minipage}}
\centerline{(b)~Comparison of performance on Separate Rooms}
\caption{Results for different maps in Overcooked environment. The first column shows the return curve in each episode~(higher is better), the second column shows the length of each episode (lower is better), and the third column shows the curve of policy entropy, demonstrating the uncertainty of the policy in action selection.}
\label{fig:overcooked}
\end{figure*}

\subsection{Performance on Overcooked}
The results on Overcooked are shown in Figure~\ref{fig:overcooked}. In the method based on raw symbolic input, CommNet showed better performance than the Symbolic-PPO. However, it is difficult to observe the communication patterns between agents in a way that humans can understand. The results also indicate that the LLM-based methods have significantly higher sample efficiency. Moreover, Our Verco further achieves higher episode returns. We believe this improvement benefits from the communication information that can effectively coordinate the actions among agents. In addition, verbal messages can also promote our understanding of the cooperation mechanisms between agents. Moreover, the results show that Verco has significantly lower episode length and entropy compared to other algorithms. This indicates that the introduction of verbal communication can also encourage agents to complete tasks more efficiently and reduce the uncertainty of action policy.
Although GPT-4 has rich prior knowledge, there are still biases in making complex decisions in the environment, which can easily lead to task failure.

\subsection{Verbal communication visualization on Overcooked}
\begin{figure*}[t]
    \centering
    \includegraphics[width=5.5 in]{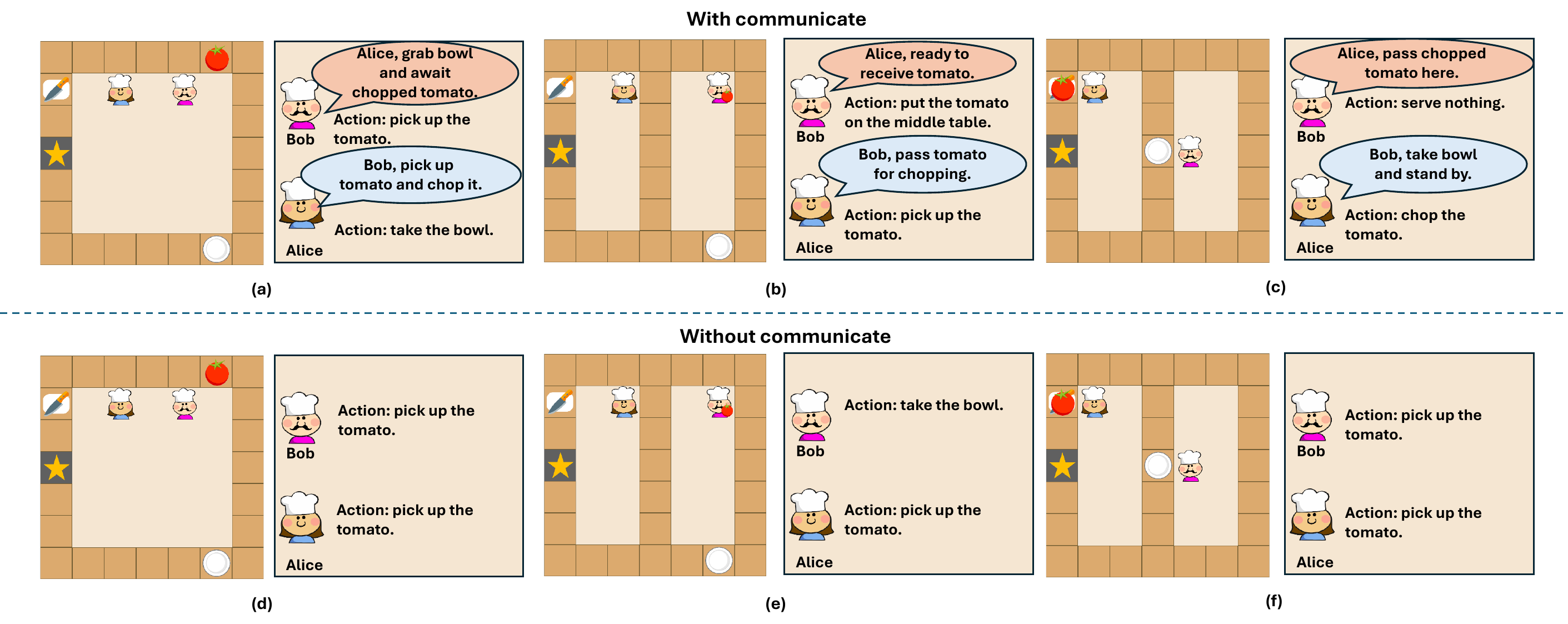}
    \caption{Communication display in different scenarios.}
    \label{replay}
\end{figure*}

To further analyze the differences brought about by the introduction of verbal communication, we visualize the replay of Verco and non-communication algorithms (TWOSOME) in detail. In the Single Room scenario as shown in Figure~\ref{replay}(a,d), communication is crucial to coordinate and perform different tasks, as there may be conflicting actions between agents. Specifically, Alice suggested that Bob should pick up the tomatoes and Bob suggested that Alice should pick up the bowl. After receiving the message from another agent, the two agents choose different actions which avoid conflicts. In the non-communication situation, both agents choose to directly pick up the tomato which will collide and waste time. In Separate Rooms, agents are separated by table and can only transfer items through the middle table. Therefore, communication is also important for agent cooperation. As shown in Figure~\ref{replay}(b), after Alice reminds Bob, Bob will put the tomatoes on the middle table for Alice to chop. Otherwise, Bob may not be aware of the need to pass the tomato to Alice without communication as shown in Figure~\ref{replay}(e). Overall, by directly sending verbal messages, we can have a more intuitive understanding of the cooperative motivations between agents.

\section{Closing Remarks}

In this paper, we propose a novel multi-agent communication algorithm, called Verco. Verco endows agents with the ability to send human-understandable verbal messages and make decisions based on teammates' messages by incorporating multiple LoRA parameters. To generate coordinated and consistent messages for the message module, we employ a Teacher LLM with global observation to produce message labels and train the message module with local observation as input. After the process of SFT, agents load well-trained communication module weights, and the action policy is trained through reinforcement learning by continuously interacting within the environment. Evaluations conducted in the Overcooked environment demonstrate that our algorithm outperforms existing baselines in terms of performance and exhibits stronger interpretability, which contributes to a deeper understanding of the formation mechanism of cooperation among agents.


\medskip

{
\small
\bibliography{ref}
\bibliographystyle{unsrtnat}
}


\end{document}